# DETOXIFYING TOXIC COMMUNICATION: A DESIGN SCIENCE APPROACH TO RESPONSIBLE AI

Hossein Arshadi Soufiani, Henry M. Kim*, Hjalmar Turesson,
Syed Mohammad Arham Noman, Anav Setia
*Schulich School of Business, York University*

## Abstract

*Toxic language in digital workplaces such as pejoratives, sarcasm, condescension, and subtle incivility can erode trust, morale, and collaboration. Existing moderation tools primarily delete or block harmful messages, disrupting communication and offering no constructive resolution. This study adopts a Design Science Research approach to create a responsible AI artifact that detects and detoxifies toxic communication. The artifact integrates fine-tuned transformer-based classifiers (DistilBERT, DistilRoBERTa) with a generative detoxification model (mT0-XL-Detox-ORPO) that rewrites toxic text into semantically equivalent, non-offensive paraphrases. Technical evaluation demonstrates high accuracy in toxicity detection and strong semantic preservation in rewritten messages, supporting conversation continuity while reinforcing respectful discourse. The paper contributes design principles for responsible AI moderation that prioritize meaning preservation and fairness.*

## 1. Introduction

Digital communication has become the backbone of collaboration in modern organizations. Team messaging platforms like Slack and Microsoft Teams are now ubiquitous, enabling instant information exchange across distributed workforces. However, along with their benefits, these platforms have also given rise to toxic communication that can undermine trust, morale, and productivity (V. K. G. Lim & Teo, 2009). Toxic language in professional settings often appears in subtle, low-intensity forms of incivility, rude or discourteous remarks that violate workplace norms of respect (Torres et al., 2024). For example, sarcasm, condescending comments, or negative gossip can pervade office chat channels without using overt slurs (Bhat et al., 2021). Such cyber incivility tends to fly under the radar because of its ambiguity, yet it can accumulate and erode organizational culture over time (Torres et al., 2024). Classic studies of workplace incivility note that these minor norm violations can spiral into more serious conflicts if left unchecked (Andersson & Pearson, 1999). Importantly, the relative anonymity and lack of face-to-face cues in online environments can exacerbate disinhibited behaviour and people may say things online that they would restrain in person. This online disinhibition effect is well-documented and especially problematic in asynchronous text channels where accountability is diffused (Suler, 2004). In response to these challenges, organizations are increasingly seeking automated solutions to detect and moderate toxic language in internal communications. Content moderation is now seen as a critical function for digital platforms rather than an afterthought. Yet, current enterprise chat moderation tools remain limited. The dominant approach is reactive removal or flagging of toxic messages, identifying a policy-violating post and deleting or quarantining them (Zhan et al., 2025). While deletion can mitigate immediate harm, it offers no constructive path to restore civility or address the underlying issue. Simply removing a toxic message may also disrupt the conversation flow and leave other participants guessing about missing context (Gillespie, 2019). This has prompted interest in more proactive and restorative moderation strategies (Gillespie, 2019). One emerging approach is text detoxification: automatically rewriting an offensive message into a semantically equivalent but non-offensive version (Logacheva et al., 2022). Rather than censoring user-generated content entirely, a detoxification system can transform a hostile message (“This idea is absolutely stupid”) into a polite rephrasing (“I have some concerns about this idea”) while preserving the original intent. Researchers posit that such toxic-to-nontoxic paraphrasing can preserve the meaning of the contribution and maintain workflow continuity, while reinforcing norms of respectful discourse (Dale et al., 2021). For instance, Dale et al. (2021) show that large pretrained language models can be fine-tuned to rephrase toxic comments into civil alternatives with minimal loss of content meaning. This detoxification technique has gained traction as a way to enhance digital discourse without erasing the speaker's voice (Dale et al., 2021).

*Corresponding author. hkim@schulich.yorku.ca
*An updated version of this paper appears in the Proceedings of the 60th Hawaii International Conference on Systems Science (HICSS-60), Honolulu, HI, January 5-8, 2027.*

To the best of our knowledge, no prior work has developed or evaluated a Design Science Research artifact that performs automatic detoxification for communication contexts while explicitly operationalizing responsible AI principles. Although existing toxicity and detoxification studies rely primarily on public social media datasets, they rarely address how detoxification should be designed for workplace communication needs such as semantic preservation, tone stability, and conversation continuity. Prior moderation tools also emphasize detection, deletion, or flagging rather than responsible transformation. This study therefore introduces a novel socio-technical artifact that integrates transformer-based toxicity detection with controlled generative detoxification to support ethical, coherent, and safe communication in organizational settings.

To address the challenges mentioned, this paper adopts a Design Science Research (DSR) approach to develop a responsible AI artifact for moderating toxic language in organizational communication channels. The remainder of the paper is structured as follows. Section 2 reviews prior work on toxicity detection, detoxification, and responsible AI in organizational settings. Section 3 describes the DSR methodology and presents the design and development of the artifact, including its system architecture and technical evaluation. Section 4 offers a detailed discussion of the artifact's implications for IS research and organizational practice, highlighting ethical considerations and design trade-offs. Section 5 offers concluding remarks, limitations and outlines directions for future research.

## 2. Literature Review

Despite progress in both automated toxicity detection and detoxification, significant research gaps remain. Many state-of-the-art toxic language classifiers struggle with nuance and context-sensitivity (Pavlopoulos et al., 2020). Toxicity is not a binary trait; what is considered unacceptable can depend on subtle cues, conversation history, and organizational culture (Zhan et al., 2025). Off-the-shelf models often use narrow definitions of abuse (e.g. focusing only on insults or hate speech) and can miss context-dependent incivility (Zhan et al., 2025). Moreover, fairness and explainability have emerged as chief concerns in recent studies of AI moderation (Mathew et al., 2021). Biased training data can lead models to systematically mislabel language from certain demographics as "toxic" (Mathew et al., 2021). For example, researchers found that some toxicity annotators were more likely to rate African American English dialect as offensive, reflecting cultural biases that can carry over into models (Sap et al., 2022). Such biases mean that marginalized communities might be unfairly targeted by automated filters. Likewise, most AI moderation systems operate as black boxes, giving end-users little insight into why a message was flagged or altered (Zhao et al., 2024). In organizational contexts, these limitations are especially pronounced, as enterprises require moderation tools that can handle domain-specific language, maintain a professional tone, and ensure fair treatment for all employees.

From a technical standpoint, traditional machine learning approaches to toxicity detection have shown both promise and limitations (Davidson et al., 2017). Supervised classifiers (e.g., variants of BERT or other transformer models) trained on labelled datasets form the backbone of most language comment detection systems (Liu et al., 2019). These models can achieve high accuracy on benchmark datasets such as Wikipedia comments or Twitter hate speech corpora (Liu et al., 2019). However, data imbalance and construct coverage remain problematic. Many curated datasets are skewed towards safe content, with toxic examples being relatively rare (often <10%) (Vidgen & Derczynski, 2021). This class imbalance can bias models towards always predicting "neutral," missing a lot of genuinely toxic incidents (low recall) (Vidgen & Derczynski, 2021). Another recent development is the application of Large Language Models (LLMs) to content moderation tasks. Within the past few years, researchers have started leveraging generative models like GPT-3/4 for zero-shot or few-shot toxicity classification (Huang, 2025). Initial studies report that LLMs can understand nuanced language and context better than earlier classifiers, yielding improvements in moderating difficult cases (Huang, 2025).

A complementary body of Natural Language Processing research focuses on generative detoxification as a constructive alternative to message deletion. Work by Logacheva et al. (2022) shows that sequence-to-sequence transformers can successfully rewrite toxic text into semantically equivalent, non-toxic phrasing, aided by beam search, semantic scoring, and re-ranking techniques. Others characterize detoxification as a constrained form of text style transfer, where preserving the original intent dramatically limits the allowable space of paraphrase (Krishna et al., 2025). Dale et al. (2021) reports that this strategy maintains conversational continuity more effectively than removal or masking, underscoring detoxification as a pragmatic method to sustain workflow without erasing communicative contributions. These studies provide empirical justification for systems that prioritize revision over censorship in digital environments (Krishna et al., 2025).

For example, researchers found GPT-based classifiers to be highly accurate in identifying various forms of harmful content across different contexts (Gilardi et al., 2023). Even a very accurate LLM might make mistakes (e.g., false accusations of toxicity), and without proper justification or appeal mechanisms, users could perceive the system as illegitimate or biased. Furthermore, the computational cost and latency of large models pose practical challenges for real-time use in high-volume communication channels (Zhan et al., 2025). Recent research has thus began examining smaller, domain-adapted models and hybrid human–AI workflows that can provide more efficient and context-aware moderation (Zhan et al., 2025). The overarching implication is that effective organizational content moderation requires more than just detection accuracy that calls for nuanced, context-sensitive systems that align with the organization's norms and values.

Beyond detecting harmful language, recent Information Systems research emphasizes that moderation mechanisms must also maintain user trust and be perceived as legitimate within organizational contexts (Glikson & Woolley, 2020). Research demonstrates that trust in algorithmic decision-making is shaped by transparency, consistency, and the perception that automated systems reflect established social norms rather than imposing arbitrary constraints (Glikson & Woolley, 2020). Similar research finds that users often react negatively to opaque removals or hidden filtering, interpreting them as censorship or surveillance rather than assistance (Jhaver et al., 2019). These findings suggest that transformation-based interventions where the meaning of a message is preserved rather than deleted may offer a more socially acceptable approach to moderation, aligning with the principles of responsible and human-centered AI proposed by (Floridi et al., 2018). The aim is to retain the substance of the original communication (so that information is not lost) while removing profanity, slurs, or insulting tone (Dale et al., 2021).

Detoxification can be viewed as a specialized case of text style transfer, where the style dimension is toxicity. Unlike generic style transfer (e.g., converting formal text to casual tone) or sentiment manipulation, detoxification imposes particularly stringent requirements on meaning preservation (Chen et al., 2022). The “content” of the message such as factual details, intent, request, etc., should remain intact after transformation. A key gap in existing detoxification research is the lack of guidance on how such systems should be designed for organizational communication, where meaning preservation and user acceptance are critical. Using a Design Science Research approach addresses this gap by providing a structured way to translate these organizational needs into concrete design requirements and an implementable, responsible AI artifact.

## 3. Artifact Design Development

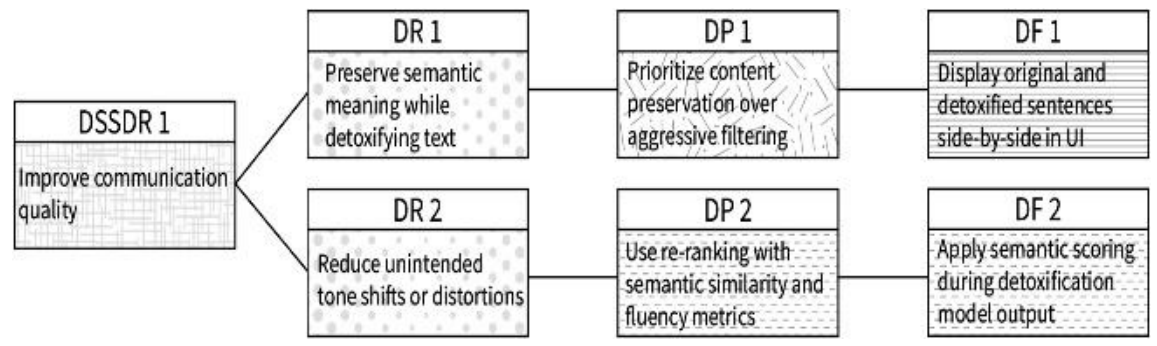

**Figure 1. Design Science Solution Design Requirements (DSSDR) Framework for Detoxification**

This study employs the Design Science Research (DSR) methodology to design and evaluate an AI-based artifact that detects and detoxifies toxic organizational communication. The approach aligns with calls for Responsible AI that address both technical effectiveness and ethical use in sociotechnical contexts (Abbasi et al., 2024; Gregor & Hevner, 2013). In DSR, the goal is to extend organizational capabilities through the creation of purposeful artifacts, constructs, models, methods, or instantiations (Hevner et al., 2004).

Our work contributes to this goal by developing a modular AI pipeline that combines toxic language classification and detoxification using large language models (LLMs). Rather than removing harmful content, the system generates non-toxic paraphrases, helping maintain communication flow and civility in professional discussions. We follow the Design Science Research Methodology (DSRM), which comprises six iterative activities: problem identification, objective definition, design and development, demonstration, evaluation, and communication (Peffers et al., 2007). These stages were implemented reflexively, with insights from each iteration guiding refinement of model configuration and dataset adaptation. The problem identification stage addressed the growing concern about subtle toxicity in workplace communication, where sarcasm or condescension can undermine team morale but remain undetected by generic moderation filters (V. K. G. Lim & Teo, 2009; Vidgen & Derczynski, 2020). The objective definition stage establishes three priorities: (1) accurate toxicity detection in contexts, (2) semantically faithful detoxification of harmful text, and (3) practical modularity for further experimentation.

Improving communication quality in organizational exchanges requires both preserving the substantive meaning of a message and moderating its interpersonal tone, as emphasized in work on cyber incivility and discourse quality in computer-mediated communication (V. K. Lim & Teo, 2009). These two requirements, in turn, motivate our design principles of content preservation and tone consistency, which we instantiate via concrete features such as side-by-side original and detoxified text, semantic similarity scoring, and re-ranking of candidate paraphrases, in line with recent text detoxification approaches (Hartvigsen et al., 2022). This refinement illustrates the iterative process of translating high-level goals into implementable design elements grounded in theory and practical utility.

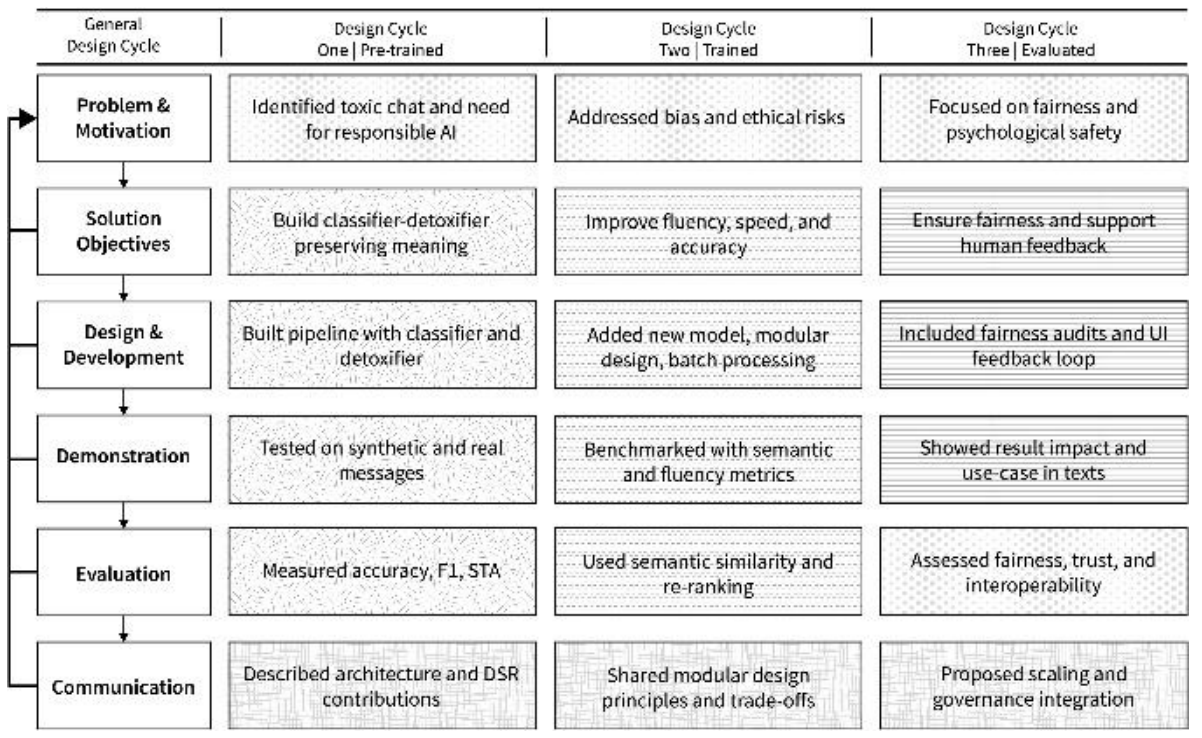

***Figure 2. Design Science Research (DSR) process across three iterative design cycles***

Given our requirements and design principles, we instantiated two modules: a toxicity classification module that decides when intervention is required and a generative detoxification module that rewrites flagged content while preserving meaning. This modular "detect-then-mitigate" architecture mirrors recent toxicity-mitigation pipelines, where a classifier first provides a toxicity signal and downstream components then reduce or reshape harmful content while maintaining utility (Cheruvu et al., 2025; Ghosh et al., 2025). By separating detection from detoxification, we can tune thresholds and models for each stage while still treating the overall pipeline as a direct instantiation of our design logic.

**Toxicity Classification Module** based on fine-tuned transformer models *DistilBERT* and *DistilRoBERTa*, trained on labeled text data to identify toxic expressions with improved contextual sensitivity.

**Detoxification Module** implemented using *mT0-XL-Detox-ORPO*, a sequence-to-sequence language model that rewrites toxic inputs into non-toxic equivalents while preserving original meaning (Logacheva et al., 2022).

**Table 1. System operation examples: toxic inputs, detoxified outputs, and changes.**

| Original message | Detoxified message | Change note |
|---|---|---|
| "Did you even read the requirements? This is a mess and a waste of everyone's time." | "I'm not sure the current version fully aligns with the requirements; could you revise it so it better addresses them?" | Preserves the core task (align with requirements) while removing personal criticism and global negative judgment. |
| "Sure, let's ignore the data again like we did last quarter 🙄." | "I'm concerned we may be underusing last quarter's data in this decision; can we revisit those results?" | Keeps the substantive critique about data use but removes sarcasm and emoji, making the concern explicit and constructive. |
| "If you can't meet deadlines, maybe this project is too much for you." | "If the current deadlines are challenging, let's discuss how we might adjust the scope or provide additional support to meet them." | Maintains the issue of missed deadlines yet removes the personal attack, reframing it as a collaborative problem-solving conversation. |

## 3.1. Artifact Design and System Architecture

The architecture establishes a modular AI pipeline designed to detect and detoxify toxic communication in organizational contexts. The detect-and-detox pipeline uses an offensive language dataset adapted from Davidson et al. (2017), who assembled a corpus of 24,802 tweets by first querying the Twitter API with a Hatebase-derived lexicon and then employing crowd annotators to assign each tweet to one of three categories: hate speech, offensive language, or neither (Davidson et al., 2017). Each tweet was labeled by at least three annotators with a 92% intercoder agreement, and the resulting distribution, 5% hate speech, 76% offensive language, and 16.6% neither, reflects the prevalence of general offensiveness relative to explicit hate (Davidson et al., 2017). This fine-grained, widely used dataset provides high-quality human judgments and clear distinctions between harmful and non-harmful language, making it suitable for adaptation to the artifact's objectives. To align with the system design, these multiclass annotations were consolidated into a binary toxic versus non-toxic scheme before splitting the data into training (80%), validation (10%), and testing (10%) subsets to support balanced model development and evaluation.

The next stage involved fine-tuning pre-trained transformer models —DistilBERT and DistilRoBERTa— on this adapted dataset. These models were selected not only for their contextual understanding of conversational language but also for their efficiency in resource-constrained settings, reflecting our intended deployment context: small and medium-sized organizations (SMOs) that lack the dedicated GPU infrastructure and machine-learning operations capacity of large enterprises. For such organizations, moderation tooling that demands extended training runs or heavyweight inference is impractical, as they cannot dedicate computing to processes running for hours or days. As knowledge-distilled variants, DistilBERT and DistilRoBERTa retain much of the contextual sensitivity of their full-sized counterparts while substantially reducing parameter count, memory footprint, and inference time, lowering the barrier to responsible AI moderation for under-resourced teams. These models were also selected for their efficiency and contextual understanding of conversational language. Fine-tuning allowed them to better detect subtle toxicity signals typical in professional communication while reducing false positives that could arise from sarcasm or informal tone. The output of this stage is a fine-tuned toxicity detection model capable of classifying text as toxic or non-toxic with high precision. The classifiers were fine-tuned for 3 training epochs using a learning rate of 5e-5 and a weight decay of 0.01, leveraging pretrained DistilBERT and DistilRoBERTa architectures with stratified sampling (random seed = 42) applied to the Davidson et al. dataset, and the full fine-tuning process required approximately 30 minutes on a T4 GPU instance in Google Colab.

Toxic sentences identified through the classifier are then passed to a detoxification model based on mT0-XL-Detox-ORPO, a sequence-to-sequence generative transformer trained to rephrase offensive sentences into semantically equivalent, non-toxic alternatives. This model performs targeted word replacement and paraphrasing while preserving fluency and contextual integrity (Davidson et al., 2017). The detoxified output is compared with the original using semantic similarity and fluency metrics to ensure that meaning is retained and that stylistic distortions are minimized. This approach aligns with Design Science principles emphasizing both functional efficacy and ethical responsibility in artifact construction.

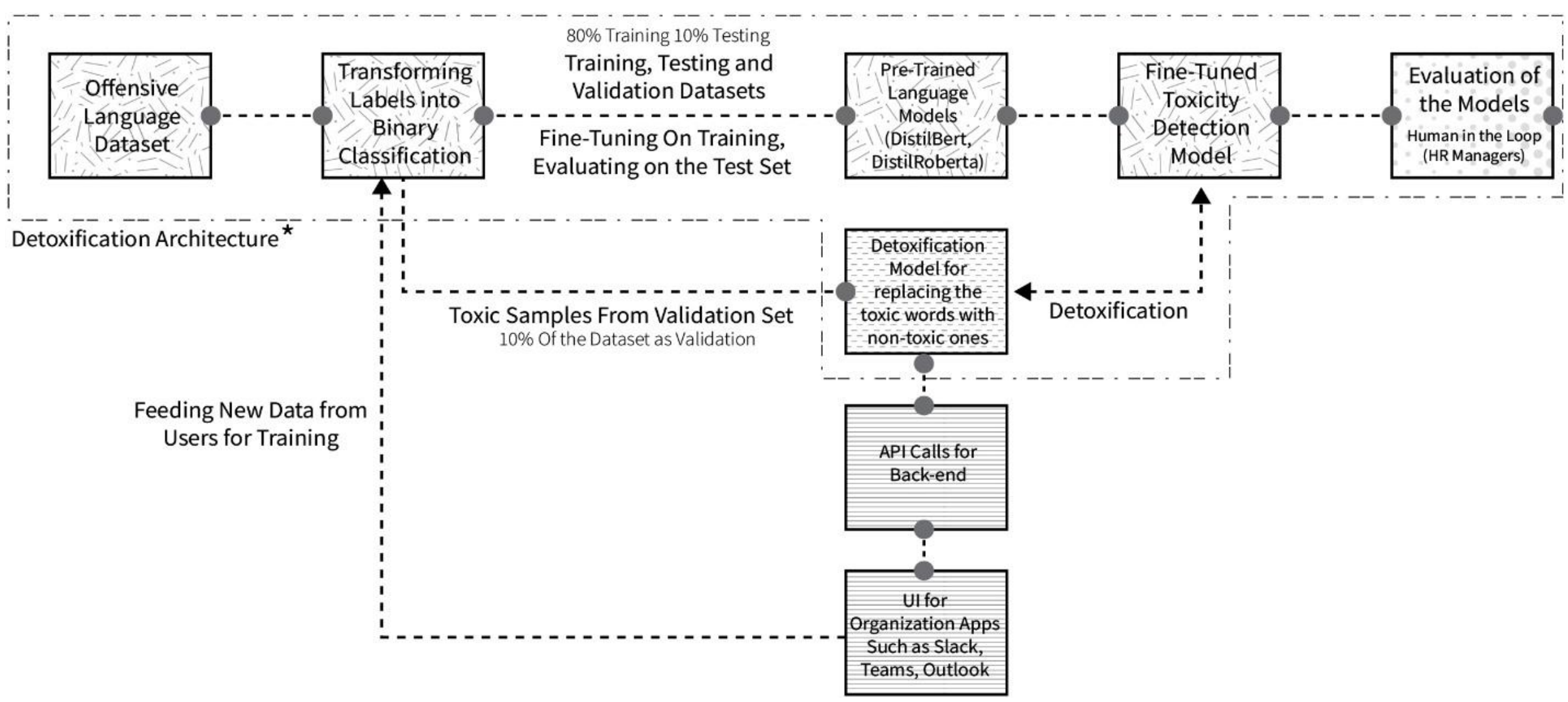

**Figure 3. Architecture of the Detoxification Artifact**

Some components in the architecture represent planned but unimplemented extensions. These include an API integration layer to enable seamless deployment within enterprise communication tools such as Slack, Microsoft Teams, and Outlook, and a user feedback mechanism designed to continuously refine model performance through active learning from new organizational data. These modules were conceptualized but not yet realized in the present artifact.

Overall, the architecture demonstrates a systematic, theory-driven approach to responsible AI system design. It integrates data preprocessing, supervised fine-tuning, and generative detoxification into a cohesive pipeline that culminates with evaluation. While real-time interaction and feedback components remain future work, the implemented system provides a reproducible foundation for scalable, ethical, and semantically aware language moderation in organizational communication.

### 3.2. Evaluation

Consistent with the Framework for Evaluation in Design Science (FEDS), the artifact evaluation was structured as a series of sequenced, low-risk assessment episodes designed to ensure both rigor and relevance (Venable et al., 2016). In FEDS terms, Episode 1 constitutes an ex-post, artificial evaluation that focuses on the artifact's technical performance, including toxicity classification accuracy and detoxification quality. Episode 2 constitutes a conceptualized ex-ante, naturalistic evaluation plan that outlines how the artifact would be deployed and assessed within an organizational communication setting. This dual orientation reflects the Design Science Research principle that evaluation is both a means of validating artifact utility and a process for generating design-relevant knowledge through iterative testing and reflection (Venable et al., 2016).

### 3.3. Technical Performance

The first evaluation episode focused on assessing the artifact's technical performance in detecting and detoxifying toxic communication while preserving semantic fidelity. The fine-tuned transformer-based classifiers, DistilBERT and DistilRoBERTa, were trained on the adapted Davidson et al. (2017) dataset, achieving strong results, accuracy of 0.965 and F1-score of 0.979, demonstrating the system's effectiveness in distinguishing between toxic and non-toxic content within conversational data.

| | Classification Model | |
|---|---|---|
| Evaluation Metrics | DistilBert | DistilRoberta |
| eval_accuracy | *0.965* | *0.960* |
| f1_score | *0.979* | *0.976* |
| precision_score | *0.980* | *0.977* |
| recall_score | *0.978* | *0.974* |

| Semantic Preservation PBert, RBert, FBert | |
|---|---|
| f1_score | 0.930 |
| precision_score | 0.969 |
| recall_score | 0.948 |

**Figure 4. Evaluation Metrics for Classification Model and Semantic Preservation Scores**

To evaluate semantic preservation, the study employed the BERTScore framework (Zhan et al., 2025), which measures token-level similarity between the original and detoxified sentences using contextual embeddings from pre-trained BERT models. This indicates that BERTScore evaluates meaning preservation by comparing the contextual embeddings of the original and detoxified sentences, allowing the model to quantify how closely the rewritten output retains the intent of the input. The detoxified outputs achieved a mean F_BERT score of 0.930, with precision (P_BERT) = 0.969 and recall (R_BERT) = 0.948, indicating that the model consistently retained the intended meaning of the input messages while successfully removing toxic elements. Most detoxified outputs scored above 0.9 on the semantic similarity scale, confirming that the model produces coherent and contextually aligned paraphrases suitable for organizational communication environments.

In addition, Style Transfer Accuracy (STA) was employed as a complementary metric (Logacheva et al., 2022) to verify that detoxified texts were recognized as non-toxic by an independent classifier. STA provides a reference-free, scalable method for validating the stylistic success of toxicity removal, supporting the overall reliability of the detoxification process. Together, these metrics confirm that the artifact not only performs effectively in toxicity classification and detoxification tasks but also upholds the broader design principle of responsible AI behavior through transformation rather than censorship, ensuring interpretability, fairness, and communication clarity in enterprise systems.

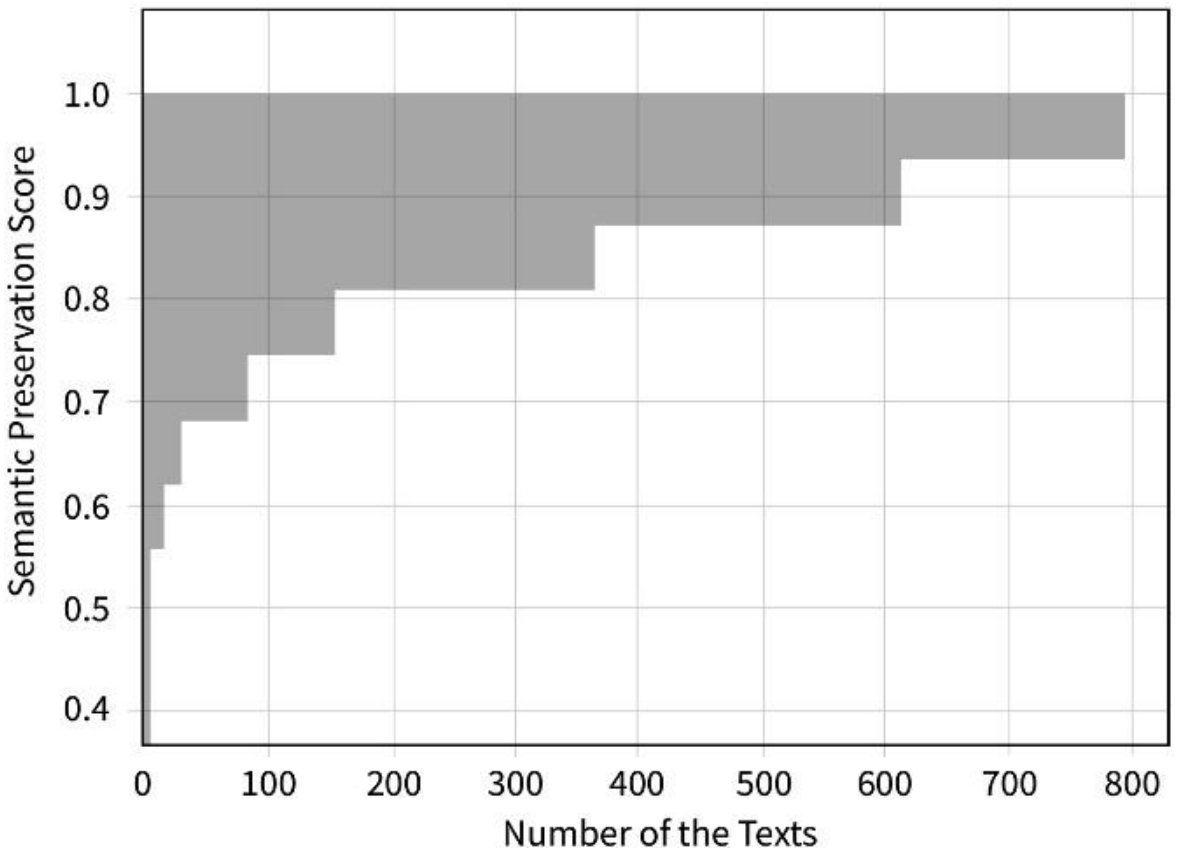

$$\text{STA} = \frac{\text{\# detoxified sentences classified as non-toxic}}{\text{\# input toxic sentences}}$$

$$\textit{[Validation Set]} \quad \text{STA} = \frac{1469}{2062} = 71.2\,\%$$

**Figure 5. Cumulative distribution of semantic and STA Score**

To support transparency and reproducibility, the full implementation including training scripts, model configurations, dataset and evaluation notebooks, is available in our public GitHub repository [https://github.com/thearhamn/detox_org_com].

### 3.4. Organizational Implementation

The second evaluation episode, organizational implementation, is conceptually designed for future deployment and testing. It involves three sequential steps:

1. Establishing a baseline for toxicity within an organizational communication environment without any intervention.
2. Introducing the detoxification intervention by implementing the full moderation pipeline in Withdrawing the intervention to measure rebound effects and assess the persistence of behavioral and linguistic improvements.

Although this phase has not yet been executed, it provides a structured roadmap for longitudinal evaluation consistent with FEDS' focus on ecological validity and real-world relevance. Future work will integrate the artifact into enterprise platforms (e.g., Slack or Microsoft Teams) via API connectors and collect user feedback to examine trust, fairness perception, and behavioral adaptation.

From a theoretical standpoint, this design reflects the principles, emphasizing that mature design knowledge should explain *why* and *under what conditions* a given artifact succeeds (Gregor & Jones, 2007). The framework ensures that the artifact's operationalization extends beyond technical accuracy to include socio-technical considerations such as fairness, ethical design, and organizational trust. In line with Gregor and Hevner (2013), this evaluative structure illustrates how iterative testing bridges practical artifact utility with theoretical generalizability.

Ultimately, this evaluation demonstrates that AI-powered artifacts in organizational contexts must balance technical robustness and human-centered design. By embedding responsible AI principles throughout its construction and validation, the artifact contributes to advancing Design Science Research in the age of autonomous, adaptive, and socially consequential AI systems.

## 4. Discussion

This study provides key implications for both IS research and practice. The developed artifact demonstrates how responsible AI principles can be embedded into digital communication tools, offering a socio-technical solution that integrates toxic language detection and detoxification. Practically, it helps organizations enhance communication ethics without losing clarity or disrupting workflows. Iterative refinement during the DSR process addressed early limitations in semantic preservation and fairness through model tuning and user feedback. However, several challenges emerged during development. Detoxification models exhibited reduced performance in multilingual or informal environments, and fairness auditing was constrained by the lack of demographic data in enterprise settings. However, existing detoxification benchmarks lack dialectal and identity-based diversity, limiting fairness evaluations (Zhou et al., 2021). Two key limitations involved the computational complexity of multilingual deployment and the availability of high-quality, demographically diverse datasets, a concern echoed in recent work documenting data selection bias and cultural narrowness in toxicity models (Sap et al., 2022). Technical trade-offs between fluency, meaning preservation, and transparency also demand careful balancing. Ethical considerations were central to the artifact's conception, emphasizing transformation rather than censorship and offering rationale-driven outputs to preserve user trust. Practical deployment of such systems also requires organizational readiness, clear communication policies, and stakeholder training to ensure effective and responsible integration. Finally, another limitation involved the computational resources available during development; we were constrained to using lighter model variants to ensure successful execution. It is plausible that access to more robust computing power could yield even better performance outcomes.

## 5. Concluding Remarks

This research presented a Design Science Research (DSR) approach to developing a responsible AI artifact for detecting and detoxifying toxic language in organizational communication. By integrating large language models into a modular framework, the study contributes to both the theoretical advancement of DSR and the practical realization of ethical AI systems. The artifact demonstrates that toxic discourse can be constructively transformed while preserving semantic meaning and conversational continuity, an essential goal for fostering civility and psychological safety in digital workplaces. The study extends the scope of traditional DSR by embedding social and ethical dimensions into the artifact design process, emphasizing transparency, fairness, and responsible transformation over censorship. The evaluation results show promising performance in semantic preservation and toxicity reduction, highlighting the potential of AI-driven moderation tools to enhance communication integrity and trust within organizations.

However, several challenges remain. Fairness auditing was limited by the absence of demographic data, and the reliance on pre-existing datasets constrained cross-cultural and contextual generalizability. The artifact's performance in multilingual and informal environments, along with computational limitations, underscores the need for continued refinement and scalability. These limitations do not diminish the artifact's contribution but rather illuminate critical design considerations for future responsible AI research in socio-technical systems.

Future work should expand the framework through human-in-the-loop evaluation, governance integration, and domain-specific deployment to capture real-world interactions and longitudinal impacts on organizational communication culture. Such extensions would strengthen the ethical maturity of AI-based interventions. Ultimately, this study illustrates that responsible AI, when guided by rigorous DSR principles, can bridge technical performance with human values. By prioritizing transformation, fairness, and meaning preservation, the proposed artifact offers a foundational step toward designing AI systems that promote constructive, inclusive, and ethically aligned digital communication.

## A. Appendix

**A1**. Below is the code snippet implementing the **Classifier_distilRoberta** model used for toxicity detection.

```
MODEL_ID = "distilroberta-base"

# Load the CSV files
train_df = pd.read_csv('train_data.csv')
val_df = pd.read_csv('val_data.csv')
test_df = pd.read_csv('test_data.csv')

# Convert pandas DataFrames to Hugging Face
   Datasets
train_dataset = Dataset.from_pandas(train_df)
val_dataset = Dataset.from_pandas(val_df)
test_dataset = Dataset.from_pandas(test_df)

# Rename 'tweet' column to 'text' for
   consistency
train_dataset =
   train_dataset.rename_column('tweet',
   'text')
val_dataset =
   val_dataset.rename_column('tweet', 'text')
test_dataset =
   test_dataset.rename_column('tweet', 'text')

# Rename 'class' column to 'labels' for
   consistency
train_dataset =
   train_dataset.rename_column('class',
   'labels')
val_dataset =
   val_dataset.rename_column('class',
   'labels')
test_dataset =
   test_dataset.rename_column('class',
   'labels')

# Preprocessing
tokenizer =
   RobertaTokenizerFast.from_pretrained(MODEL_
   ID)

def tokenize(batch):
    result = tokenizer(batch["text"],
   padding=True, truncation=True,
   max_length=256)
    result["labels"] = batch["labels"]
    return result

train_dataset = train_dataset.map(tokenize,
   batched=True,
   batch_size=len(train_dataset))
val_dataset = val_dataset.map(tokenize,
   batched=True, batch_size=len(val_dataset))
test_dataset = test_dataset.map(tokenize,
   batched=True, batch_size=len(test_dataset))

# Set dataset format
train_dataset.set_format("torch",
   columns=["input_ids", "attention_mask",
   "labels"])
val_dataset.set_format("torch",
   columns=["input_ids", "attention_mask",
   "labels"])
test_dataset.set_format("torch",
   columns=["input_ids", "attention_mask",
   "labels"])

# Define labels
num_labels = 2  # Binary classification: 0
   (not offensive) and 1 (offensive)
id2label = {0: "not_offensive", 1:
   "offensive"}
label2id = {"not_offensive": 0, "offensive":
   1}

print(f"number of labels: {num_labels}")
print(f"the labels:
   {list(id2label.values())}")

# Update the model's configuration
config = AutoConfig.from_pretrained(MODEL_ID)
config.update({
    "num_labels": num_labels,
    "id2label": id2label,
    "label2id": label2id
})
```

**A2**. Below is the code snippet implementing the **Detoxifier** model used for detoxification.

```
model =
   AutoModelForSeq2SeqLM.from_pretrained('s-
   nlp/mt0-xl-detox-orpo')
tokenizer = AutoTokenizer.from_pretrained('s-
   nlp/mt0-xl-detox-orpo')

LANG_PROMPTS = {
   'zh': '排毒：',
   'es': 'Desintoxicar: ',
   'ru': 'Детоксифицируй: ',
   'ar': 'إزالة السموم: ',
   'hi': 'विषहरण: ',
   'uk': 'Детоксифікуй: ',
   'de': 'Entgiften: ',
   'am': 'መርዝ መርዝ: ',
   'en': 'Detoxify: ',
}

def detoxify(text, lang, model, tokenizer):
   encodings = tokenizer(LANG_PROMPTS[lang] +
   text, return_tensors='pt')

   outputs = model.generate(**encodings,
                            max_length=128,
                            num_beams=10,

   no_repeat_ngram_size=3,

   repetition_penalty=1.2,
                            num_beam_groups=5,

   diversity_penalty=2.5,

   num_return_sequences=5,

   early_stopping=True,
                            )

   return tokenizer.batch_decode(outputs,
   skip_special_tokens=True)
```